\documentclass[a4paper, amsfonts, amssymb, amsmath, reprint, showkeys, nofootinbib, twocolumn]{revtex4-2}
\usepackage[english]{babel}
\usepackage[utf8]{inputenc}

\usepackage{amsthm}
\usepackage{mathtools}
\usepackage{xcolor}
\usepackage{graphicx}
\usepackage[left=23mm,right=13mm,top=35mm,columnsep=15pt]{geometry} 
\usepackage{adjustbox}
\usepackage{placeins}
\usepackage[T1]{fontenc}
\usepackage{lipsum}
\usepackage{csquotes}
\usepackage{lineno}
\usepackage{soul}
\usepackage{amsmath}
\usepackage{amsfonts} 
\usepackage{listings}

\begin{document}

\title{Application of machine learning technique for a fast forecast of aggregation kinetics in space-inhomogeneous systems}
\author{M.A. Larchenko}\email[Corresponding author:  ]{mariia.larchenko@gmail.com}

\author{R.R. Zagidullin,
         V.V. Palyulin, 
         N.V. Brilliantov}
\affiliation{Skolkovo Institute of Science and Technology, Bolshoy Boulevard 30, Moscow, 121205, Russia}


\begin{abstract}
    Modeling of aggregation processes in space-inhomogeneous systems is extremely numerically challenging since  complicated aggregation equations -- Smoluchowski equations are to be solved at each space point along with the computation of particle propagation. Low rank approximation for the aggregation kernels can significantly speed up the solution of Smoluchowski equations, while particle propagation could be done in parallel.  Yet the simulations with many aggregate sizes remain quite resource-demanding. Here, we explore the way to reduce the amount of direct computations with the use of modern machine learning (ML) techniques. Namely, we propose to replace the actual numerical solution of the Smoluchowki equations with the respective density transformations learned with the application of the conditional normalising flow. We demonstrate that the ML predictions for the space distribution of aggregates and their size distribution requires drastically less computation time and agrees fairly well with the results of direct numerical simulations. Such an opportunity of a quick forecast of space-dependent particle size distribution could be important in practice, especially for the online prediction and visualisation of pollution processes, providing a tool with a reasonable tradeoff between the prediction accuracy and the computational time.
\end{abstract}

\maketitle

\section{Introduction}

Particle aggregation is a widely spread process where small particles merge and form larger clusters or aggregates \cite{Krapivsky,Leyvraz}. In atmospheric processes the airborne particulates coalesce into smog particles \cite{Srivastava1982}, while aggregation of water aerosols forms cloud droplets or ice crystals. These   phenomena effect in cloud formation and precipitation, e.g. \cite{Friedlander,Klett,Seinfeld,Falkovich2002}. An aggregation of hard fractions suspended in the atmosphere influences their sedimentation rates and, thus, affects the  air quality; similar processes could happen in liquid media \cite{brilliantov2023aggregation}.

Air pollutants can originate from a range of sources, like industrial facilities, city traffic, agriculture, coal combustion and wildfires \cite{Friedlander}. According to some studies, both the size and the origin of pollutants are important for public health \cite{Zhang_Boya_Weuve2023, Clifford2016}. PM2.5 and PM10 are particles with diameters smaller than 2.5$\mu m$ and 10$\mu m$ respectively. Several investigations associate fine particulate matter pollution (PM2.5 and PM10) with a lower life expectancy \cite{bennett2019particulate} and higher risk of diseases, such as lung cancer \cite{UKreport}, incident dementia \cite{Zhang_Boya_Weuve2023} and heart failure \cite{Shah2013}. This small-sized matter can spread over long distances by diffusing and following the winds. Thus, for many practical problems one often needs to know the spatial distribution of particles of various sizes provided that the location of their source of is known (e.g. a volcano, factory chimneys, etc.) and that the particles aggregate. Such a distribution would sensitively depend on the wind direction. It may also happen that bursts of particles occur from time to time, due to natural processes or industrial cycles. That underlines the importance of modeling and prediction of transport of these air pollutants in cities and other populated areas for urban planning and monitoring when a rapid prediction of the pollution level
of the air by particulates is often needed.

Another important application of such models could be the solution of the inverse problem. That is, one may wish to detect the origin of a pollution source using the data from a system of rather sparsely spread observation stations. All these problems require a development of precise and computationally efficient models of diffusion-aggregation processes.

\section{Conventional approach: The numerical solution of space-inhomogeneous aggregation equations.}

\subsection{Space-inhomogeneous Smoluchowski equations}
Conceptually, it is not difficult to formulate a space-inhomogeneous model with aggregation, as all the comprising components are available -- the aggregation Smoluchowski equations \cite{Krapivsky,Leyvraz} and the hydrodynamic equations with the diffusion that describe the motion of particle due to advection and diffusion \cite{galkin2001}. For the case of one-dimensional (1D) model with the inhomogeneity in one direction \cite{Asymmetric:source,Kirone:source,zagidullin2022}, an analytical solutions can be found. 
Generally, one has to investigate the respective processes numerically \cite{hackbusch2012numerical, bordas2012numerical,chaudhury2014computationally,zagidullin2022}. For aggregation in space-inhomogeneous systems the kinetic equations can be obtained from the conventional Smoluchowski equations by supplementing them with the advection and diffusion terms which yields \cite{zagidullin2022}, 
\begin{equation}
    \begin{split}
    \frac{\partial c_k}{\partial t} + ({\bf V}_k \cdot  {\bf \nabla}) c_k = \frac{1}{2}\sum_{i+j=k}K_{ij}c_ic_j \\
    - c_k\sum_{j\geq 1}K_{kj}c_j + D_k \Delta c_k + I\delta_{k,1}\delta(\bf r),
    \label{loc:input}
    \end{split}
\end{equation}
where $c_k=c_k({\bf r}, t)$ denote the number densities (a number of particles per unit volume) of clusters of size $k$ at time $t$ and point ${\bf r}$. The size of a cluster $k$ is a number of elementary units or monomers in the aggregate. The second term on the l.h.s. describes advection (drift) with the advection velocity ${\bf V}_k$ for clusters of size $k$. On the r.h.s. the first term describes the kinetics of aggregation with aggregation rate coefficients,  $K_{ij}$, quantifying the reaction rates of the cluster coalescence, $[i]+[j] \to [i+j]$ (the prefactor $1/2$ prevents double counting). The second term there quantifies the reduction of $c_k({\bf r}, t)$ due to aggregation of such particles with other aggregates or monomers. The third  term on the r.h.s. of Eq. \eqref{loc:input} accounts for cluster diffusion with the diffusion coefficient $D_k$. The last term on the right-hand-side of Eq. \eqref{loc:input} describes the source of monomers located at the origin ${\bf r}=0$. Eq. \eqref{loc:input} corresponds to the Euler level of hydrodynamic description of the system. Here we assume that the rates $K_{ij}$ have the same form as for space-homogeneous systems (see Ref. \cite{brilliantov2023aggregation} for the respective generalisation).

\subsection{Numerical solution of space-inhomogeneous Smoluchowski equations}

Solving space-homogeneous Smoluchowski equations is already numerically challenging. Indeed, the system of equations \eqref{loc:input} is an infinite system of ODEs.  Hence, it should be approximated by a finite set of equations which contain cluster densities up to some value $k_{\rm max}$. If one starts with some reasonable initial conditions, with non-zero densities $c_k({\bf r},0)$ for $1\leq k \leq k_0$, where $k_0$ is small ($k_0 <5$), then larger clusters appear at later time. After time $T$ clusters of maximal size of $k_{\rm max}(T)$ emerge. In order to make simulations meaningful, one needs to use at least $k_{\rm max}(T)$ equations to model the evolution of the system till time $T$. As it is shown for some analytically treatable models, the characteristic size of clusters increases with time as $\sim t^{\alpha}$, where $\alpha$ is a positive constant, which depends on the form of the kernel $K_{ij}$. $\alpha$ can exceed unity \cite{Krapivsky,Leyvraz}. Hence, one concludes that the number of equations needed for the description of the system evolution till time $T$ scales as $k_{\rm max}(T) \sim T^{\alpha}$, i.e. it rapidly grows with $T$, making the numerical solutions of these equations computationally demanding. For the low-rank decomposition methods \cite{matveev2015, matveev2018parallel,skorych2019investigation,osinsky2020low,MKSTB,OscPRE2017} as well as for other state-of the art methods the computational complexity scales as $O((N\log N) T)$, where $N$ is the number of equations and $T$ is the number of computational time steps.

The solutions of space-nonhomogeneous aggregation equations, obviously, is more complicated, since one needs to compute local concentrations $c_k({\bf r},t)$ which means solving the system of Smoluchowski equations \eqref{loc:input} for each position. In practice we use the following approach. The first two terms on the r.h.s. of these equations, the coagulation operators, were calculated explicitly, while for the diffusion and advection operators we use the standard representation as suggested in Ref. \cite{Vabishchevich}. The monomer source was implemented through the boundary conditions of the second type which read (see also \cite{zagidullin2022} for 1D case), 
\begin{eqnarray}
    \label{BC}
    &&c_1' (x=+0,y=+0,t)= -I/4; \\
    &&c_k'(x=+0,y=+0,t)= 0 \qquad {\rm for } \qquad k\geq 2. 
\end{eqnarray}
To place a monomer source at the boundary of one of the axes, the boundary condition for monomers is modified using the Gaussian function, e.g:
\begin{eqnarray}
    \label{BC_wall}
    &&c_1' (x=+0,y=y_s,t)= -I/2 \cdot e^{-(y-y_s)^2/\sigma_y^2} \, ,
\end{eqnarray}
where $y_s$ is the $y$-coordinate of the source of monomers, $\sigma_y^2$ is the dispersion of the source along the $y$-axis.

To find the solution numerically we use the operator-splitting method, i.e.~we solve for the problem coagulation, advection and diffusion operators separately. This allows us to incorporate efficient solvers for each process without making the numerical scheme too cumbersome. Diffusion equation is solved with the finite difference approach. We use the implicit scheme as doing otherwise would impose a very restrictive Courant condition. Advection solver uses the Lagrangian approach. This allows us to evade the artificial diffusion that comes from the Eulerian-based solutions. The coagulation step is implemented in the form of a finite difference scheme. We also apply the low-rank approximation for the coagulation kernel, see e.g. \cite{matveev2015, matveev2018anderson, skorych2019investigation, osinsky2020low} for details. The coagulation kernel in this approximation can be is written as,
\begin{equation}
    K_{ij}=K(i,j) = \sum_{\alpha=1}^{R} a_{\alpha}(i)b_{\alpha}(j) \, ,
    \label{cross_algo}
\end{equation}
where $R$ is the rank of decomposition. For the constant kernel, $K_{ij}=1$  or product kernel, $K_{ij}=(ij)^{\nu}$, the corresponding components are $a_{\alpha}=b_{\alpha}=1$,  or $a_{\alpha}=i^{\nu}$ and $b_{\alpha}=j^{\nu}$; obviously, $R=1$ here.  With the approximation \eqref{cross_algo} and for small $R$, the Smoluchowski equations can be solved  much faster.  If the ranks of the considered kernels are equal to $R$, one needs $O(N \log N R)$ operations for each time step for a spatially uniform system, where $N$ is the number of particle size grid points and  $O(M N \log N R)$ for a nonuniform one, with $M$ spatial grid points  \cite{zagidullin_2017}. In other words, the application of the low-rank decomposition for $K_{i,j}$ dramatically decreases the computational time. Noteworthy, the fast algorithms can be efficiently exploited at modern computing clusters in parallel, with the use of thousands of CPU-cores and GPU accelerators \cite{matveev2018parallel,zagidullin2019supercomputer} \footnote{The full numerical solver is available at \url{github.com/RishatZagidullin/coagulation-diffusion-2d}}.

The initial densities are zero for every cluster type:
$$ c_k(x,y, t=0) = 0\, , \qquad (x,y) \in S \, ,$$
where $S$ is the computational domain, $S=[0, X_{max}] \times [0, Y_{max}]$.

Excluding the source of monomers \eqref{BC} or \eqref{BC_wall}, the boundary conditions are also zero
$$
c_k((x,y) \in \delta S', t) = 0 \, , \qquad t \in [0, T_{max}]  \, ,
$$
where $\delta S'$ is the boundary of the computational domain that excludes the source of monomers.

In general, zero Dirichlet boundary conditions can lead to slight boundary reflection errors. Thus, one needs to use absorption layer techniques \cite{berenger1994}. However, for sufficiently large propagation distances the oscillation errors can be negated. During our simulations such errors did not occur.

As it follows from the discussion above, the solution of the space-nonuniform Smoluchowski equations requires significant computational resources even with the use of the most advanced techniques. As we show below, a model based on neural networks can be trained on the available solutions of the equations for some set of reference parameters and then it can predict the solutions for other parametric sets.

\subsection{Numerical generation of the dataset}

In practical application of the artificial neural networks (ANNs), one uses two datasets. The first one is used for training and the second one for checking the prediction accuracy. Therefore, we generate train and test datasets for $k_{\rm max}= 500$ using the mentioned above fast numerical solver. The dataset contains 500 images of $\log c(x, y)$ for every considered combination of advection velocity $V$ and the source intensity $I$ as well as a concentration coefficient that allows to restore absolute value of concentration from an image.



We used an adaptive simulation grid to model the aggregation process during 6 \texttt{hours} in the area of 10x10 \texttt{kms}. Initially, monomers and other particles were absent and the source was generating monomers only. The diffusion coefficient was set to $1\, m^2/s$ (which corresponds to the turbulent diffusion) and the advection velocity to $0.5\, m/s$. The radius of the monomer source was $0.1\, km$ and the intensity of the source in units of $cm^{-3}/s$ had the following values, $I_{\rm train}= 10,\, 20, \,30, \,40, \,50, \,60, \,70, \,80, \,90, \,100$. The total number of conditions (could be seen as labels) was 10 000. We tested the accuracy of the prediction on a dataset with other intensities $I_{\rm test}=5,\, 11, \,15, \,25, \,55, \,95, \,105$ which were not used during the model training.

\begin{figure}[t]
    \centering
    \includegraphics[width=\linewidth]{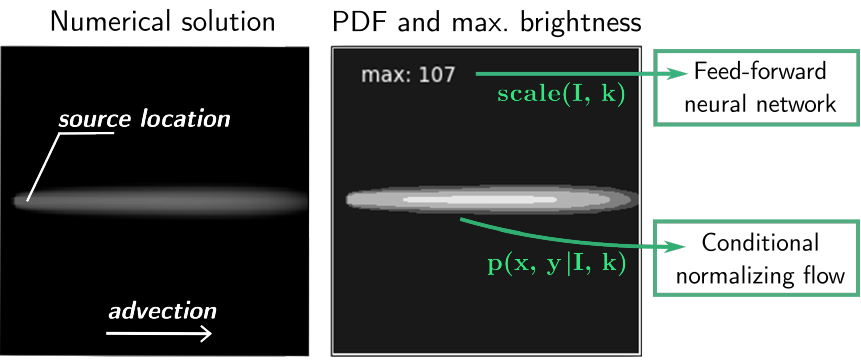}
    \caption{The model is divided into 2 modules to achieve a high accuracy. Feed-forward network is trained to approximate a maximum brightness of image $scale(I, k)$ and conditional normalising flow learns a probability density function $p(x, y|I, k)$, where $I$ is the source intensity and $k$ is a particle size.}
    \label{fig:source_model}
\end{figure}

\section{Space-inhomogeneous aggregation. Application of machine learning}

From the previous section it is clear that the straightforward simulations of pollution spread with the space-inhomogeneous Smoluchowski equations is a rather time consuming process hardly producing a rapid forecast. We show here that ANNs present a viable alternative to the traditional numerical approaches in this case. A model can be trained on data obtained from experimental observations or extensive numerical simulations. Once trained, it can produce smooth transition to the solutions under unseen initial parameters much faster than a numerical solution of equations. 

While the general forecast is a global aim of such kind of research, here we merely demonstrate the concept, restricting ourselves to a more basic problem. Namely, we focus on stationary states of a system, with a localised source of monomers. Moreover, for simplicity, we assume that the main kinetic coefficients $K_{ij}$, $V_k$ and $D_k$ are constant and do not depend on the cluster size. The generalisation for the mass-dependent kinetic coefficients is straightforward and will be addressed in the future studies. Hence, to this end, we assume $K_{ij}=K$ and $D_k=D$ to be constant and vary ${\bf V}_k= {\bf V}$ and $I$, since the latter two values quantify the environmental changes.

In the present study we train an ANN on the data obtained by a direct numerical solution of the above Eqs. \eqref{loc:input}. In this way we generate cluster densities $c_k(x,y)$ for the aggregates of size $k \in 1...k_{\rm max}$. Due to a huge difference in magnitudes of $c_k(x, y)$ for different $k$ it is convenient to store the solution as $\log c_k(x, y)$. Moreover, we convert $\log c_k(x, y)$ into integers from 0 to 255, and then use the respective scaling coefficient for restoration of the original $c_k(x, y)$ in SI units. In other words, for a given pair ${\bf V}$ and $I$, a set of $k_{\rm max}$ images, $\log c_k(x, y|\vec{V}, I, k)$, the monochrome two-dimensional patterns, can be associated. The values of $\log c_k(x, y)$ in these patterns are encoded by the brightness degree (see e.g. Fig. \ref{fig:source_model}).

For the case of a constant velocity field varying the advection velocity ${\bf V}$ can notably alter the spacial distribution of $c_k(x, y)$. Still the impact of this quantity is limited. As it follows from the basic Eqs. \eqref{loc:input}, it is sufficient to make reference simulations for a fixed (reference) value of ${\bf V}_r$, and then obtain a solution for other values of ${\bf V}$ by the respective affine transformation. The latter include rotation and re-scaling with respect to $x$ and $y$ axes. At the same time, this is not true for the intensity of a source of monomers $I$. Its variation changes the shape of $c_k(x, y)$ distribution in a non-affine way.

Keeping this in mind, we decided to divide a model which approximates the spatial distribution $c(x, y| \vec{V}, I, k)$ into several modules. Instead of learning directly the values of $\log c(x, y|\vec{V}, I, k)$, we treat every image in the dataset as a two-dimensional probability density function by normalising the original data while saving the maximum brightness of an image as a scale. To achieve the high accuracy, we use a conditional normalising flow (CNF), for learning the conditional probability density function $p(x, y|I, k)$, and additionally, a separate network, that learns the scaling factor (the maximum brightness of image) $scale(I, k)$ for the reference velocity ${\bf V}$. This is illustrated in Fig. \ref{fig:source_model}. Finally, the conditional affine transformation is to be applied to rotate and re-scale accordingly the distribution, to fit the desired advection velocity. Since the CNF-related step is the most challenging one, we present the most detailed discussion for this part of the model.

\subsection{Conditional normalising flow}

The conditional normalising flow (CNF) is a generative model that is able to learn complicated conditional distributions. CNF extends a family of normalising flows which gained significant attention in recent years for their capacity to model complex high-dimensional distributions and perform efficient sampling and density estimation \cite{kobyzev2020normalizing,papamakarios2021normalizing}.
The general idea of a normalising flow is to map a simple prior (base) distribution $p_z(z)$, such as a multivariate Gaussian, to a more complex target (data) distribution $p_x(x)$ through a series of invertible transformations $f_i$.

Training of the flow is usually done in a ``normalising'' direction, i.e. ``moving'' transformed samples $z = f(x)$ from target $x$-space closer to prior density $p_z$. Then the flow is able to produce target samples $x = f^{-1}(z)$ via an inverse function $f^{-1}$.

Assuming a bijective function $f$ such that $z = f(x)$ and performing the change of variables for the probability density gives
\begin{eqnarray}
  &&p_x(x) dx  =  p_z(z) dz, \\
  &&p_x(x)  =  p_z(f(x)) \bigg|\det\frac{df(x)}{dx}. \bigg| \label{eqn:change}
\end{eqnarray}
Here, the first term $p_x(x)$ is a likelihood we aim to maximize, $p_z(f(x))$ is a likelihood of the transformed data under the base measure, and the last term is Jacobian determinant of the transformation.

When the mapping $f$ is done by a chain of $n$ $f_i$ functions $z = f(x) = f_n(\dots f_1(x))$, it is convenient to recast Eq. \eqref{eqn:change} into a sum
\begin{equation}
    \begin{split}
  \log p_x(x) & =  \log \bigg( p_z(f(x)) \bigg|\prod_{i=1}^n\det\frac{df_i(x_i)}{dx_i}\bigg|\bigg) \\
  & = \log p_z(f(x)) + \sum_{i=1}^n \log\bigg|\det\frac{df_i(x_i)}{dx_i}.\bigg|
  \label{eqn:loglike}
  \end{split}
\end{equation}
Thus, for an efficient training and density estimation, the functions, $f_i$, have to be bijective, differentiable and their log-determinant has to be easily computed.

For a fast sampling, one should ensure that the inverse $f_i^{-1}$ is computationally tractable (see \cite{kobyzev2020normalizing} for a review and \cite{papamakarios2021normalizing} for more detail about the technical implementation of the method).

A model used in this work is based on the Real NVP (real-valued non-volume preserving) architecture by \cite{dinh2016density} with the conditioning mechanism described in \cite{winkler2019learning}\footnote{The code is available at \url{github.com/maria-larchenko/aggregation_cnf}}. The Real NVP flow is constructed as a chain of linear transforms $y = f_i(x)$ called affine coupling layers. For $x, y \in \mathbf{R^D}$ a single layer is
\begin{equation}
    \begin{split}
   y_{1:d} & = x_{1:d}, \\
   y_{d+1:D} & = x_{d+1:D} \odot \exp\big(s(x_{1:d})\big) + t(x_{1:d}),
  \end{split}
\end{equation}
where $d < D$, and "$\odot$" denotes the element-wise product. 

At the first step, affine coupling layer splits the input vector into two parts $x = [x_{1:d}, x_{d+1:D}]$. The scale $s(\cdot)$ and translation $t(\cdot)$ operations are the neural networks that take as argument the first part. Then $s(x_{1:d})$ and $t(x_{1:d})$ transform only the second part $x_{d+1:D}$ of the input. The output $y$ is a concatenation of unmodified and modified vectors $y = [y_{1:d}, y_{d+1:D}]$. The next coupling layer repeats the procedure, changing now the first part $x_{1:d}$ using the second $s(x_{d+1:D})$ and $t(x_{d+1:D})$. This technique is called an alternating pattern and allows to modify whole vector $x$ in two steps.

By adding conditioning variables $c$ into Eq.  \eqref{eqn:loglike}, one can extend capabilities of the original flow with a few changes in the training process:
\begin{equation}
  \log p_x(x|c) = \log p_z(f(x)|c) + \sum_{i=1}^n \log\bigg|\det\frac{df_i(x_i, c)}{dx_i}\bigg|
  \label{eqn:condloglike}.
\end{equation}
Conditioning for RealNVP models is introduced straightforwardly, by appending the conditioning variables $c$ to the $x_{1:d}$ part in all affine coupling layers. In terms of the model architecture, it requires only to increase a number of input units in $s(\cdot)$ and $t(\cdot)$ networks \cite{winkler2019learning},
\begin{equation}
    \begin{split}
   y_{1:d} & = x_{1:d}, \\
   y_{d+1:D} & = x_{d+1:D} \odot \exp\big(s(x_{1:d}, c) \big) + t(x_{1:d}, c). 
   \label{eqn:affine_trans}
   \end{split}
\end{equation}
Noteworthy, this approach involves the estimation and approximation of 2D density $p(x, y| c)$ in contrast to the image processing, where dimensionality of $p$ is much higher. This results in noticeable difference in the architecture of the normalising flow and its training process. The parameters of the model are given in Table \ref{tab:CNF}. Our best performed CNF model consists of 32 conditional affine transformers with 4 layers and 64 hidden units. For a two-dimensional problem, the input dimension is one and the affine coupling with two conditional units is performed, as demonstrated by Eq. \eqref{eqn:affine_trans}.

 \begin{table}
 \caption{CNF module and its training process}
 \label{tab:CNF}
 \centering
 \begin{tabular}{l c}
 \hline
   Conditional RealNVP &  \\
 \hline
   affine layers &  32  \\
   type of $s$ and $t$ nets$^{*}$ & feed-forward \\
   $s$ net layers & 4  \\
   $s$ net hidden units & 64  \\
   activation  & LeakyReLU  \\
   output regularisation  & Tanh  \\
   \hline
   loss  & negative log-likelihood \\
   optimiser  & ADAM \\
   learning rate &  $10^{-4}$ - $10^{-5}$ \\
   conditions per batch  &  4 - 5\\
   samples per condition & 512 \\
 \hline
 \multicolumn{2}{l}{$^{*}$Scale and translation nets have the similar architecture}
 \end{tabular}
 \end{table}

\subsection{CNF and FFN modules: The training process}

A standard practice of image processing implies an applying a batch normalisation procedure to ensure that inputs of linear layers do not deviate too much  \cite{ioffe2015batch}. In our case this procedure is partly replaced by the function $\operatorname{Tanh}(\cdot)$ in the CNF module. However, we still need to re-scale an input of conditional units for every conditional affine transformer of the flow. Since the model we develop is aimed only to work for a given range of $I$ and $k$, we observe that it is sufficient to divide the conditional input by approximate maximum values of $I_{\rm max} \sim 100$ and $k_{\rm max} = 500$.

The input data for the CNF module are sampled with the direct Monte-Carlo, where a target distribution $p(x, y|I, k)$ is given by a data image for $x, y, \in [-0.5, 0.5]$ domain. The prior density $p_z(x, y)$ is a multivariate normal with $\mu_x, \mu_y = 0$ and $\sigma_{xx}, \sigma_{yy} = 1$.

As usual we train the conditional normalising flow with negative log-likelihood, given by Eq. \eqref{eqn:loglike}. For RealNVP flows the Jacobian of a single layer $f_i(x)$ has a triangular form \cite{dinh2016density}
\begin{equation}
\frac{\partial f_i(x)}{\partial x^T}=\left[\begin{array}{cc}
\mathbb{I}_d & 0 \\
\frac{\partial y_{d+1: D}}{\partial x_{1: d}^T} & \operatorname{diag}(\exp (s))
\end{array}\right],
\end{equation}
where $\operatorname{diag}(\exp (s))$ is the diagonal matrix with diagonal elements corresponding the scales,
thus the Jacobian determinant is a sum of the $s_j$ scales.

For $n$ layers of the 2D CNF module $f(x, y| I, k)$ its loss function is a negative log-likelihood
\begin{equation}
  \begin{split}
  \operatorname{loss_{~CNF~}}(I, k) = &- \mathbb{E}_{x, y}\Big[\log p_z(f(x, y| I, k)) \\
   & + \sum_{i=1}^n s_i(x_i, y_i| I, k) \Big],
  \end{split}
  \label{eqn:cnf_loss}
\end{equation}
where $p_z$ is the density of the transformed sample under Gaussian base measure. Since a batch is constructed for $N$ randomly chosen conditional pairs, a total loss is averaged over the pairs $(I, k)$.

Turn now to the training of FFN module. It is much simpler, as may be seen from Table \ref{tab:FFN}. Its input is only a conditional pair $(I, k)$ and its label is maximum brightness level of the corresponding data image. Architecture of the module is almost the same with the scale and translation nets of the affine transformers. The initial output of this module bounded by $[-1, 1]$ is scaled by 255. For the training we use Huber loss.

\begin{figure}[t]
    \centering
    \includegraphics[width=\linewidth]{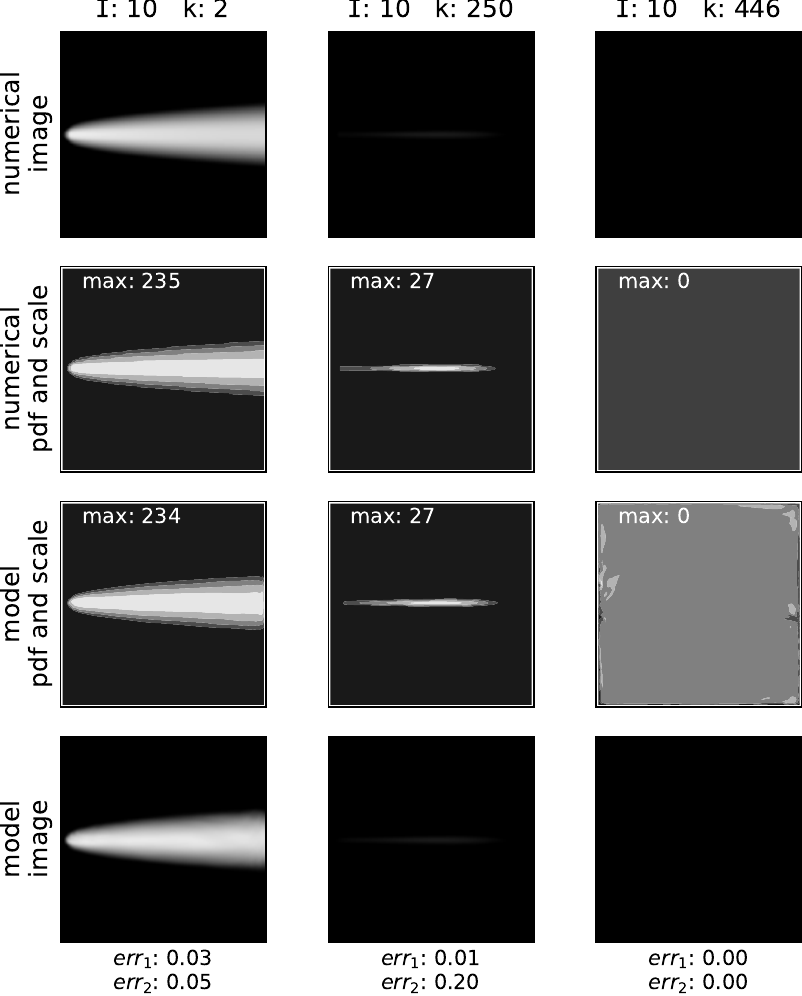}
    \caption{Restoration of data on which the model was trained. The top row is the original images, the second row is the ``technical'' representations of the images as a course-grained $p(x, y| I, k)$ and brightness levels. The thirds row is the same representations generated with the CNF and FFN modules, the bottom row represents the reconstructed images. The mean relative error is $\sim5\%$.}
    \label{fig:restoring_im}
\end{figure}

 \begin{table}[t]
 \caption{FFN module and its training process}
 \label{tab:FFN}
 \centering
 \begin{tabular}{l c}
 \hline
   Feed-Forward Network $scale(I, k)$ &  \\
 \hline
   layers & 4  \\
   hidden units & 64  \\
   activation  & LeakyReLU  \\
   output regularisation  & $255 \ \cdot$ Tanh  \\
   \hline
   optimizer  & ADAM \\
   learning rate &  $10^{-4}$ \\
   loss  & Huber loss \\
   batch size &  128 \\
 \hline
 \end{tabular}
 \end{table}

\section{Results} 

\subsection{Restoration of images and prediction of unseen data}

Fig. \ref{fig:restoring_im} demonstrates the procedure of the reconstruction of the data images used in the training (that is, seen by the model). Normalising flow is trained with samples from the desired conditional distribution and is able to perform a direct density estimation of $\log p(x, y| I, k)$ with \texttt{cnf\_pdf} up to a constant, so the restoration of image contains additionally the pdf re-scaling, which changes scale to the brightness level, learned separately by the FFN module.
%
%

We calculate two types of errors to assess the accuracy of the results. The first one, ${\rm err_1}$, is pixel-to-pixel comparison of true and reconstructed images, divided by the absolute value of the true image, see Eq. \eqref{eq:err_1}. Note, that ${\rm err_1}$ shows the mean relative error only for non-zero pixels of the true image, otherwise it is undefined. 
The second one, ${\rm err_2}$, defined by Eq. \eqref{eq:err_2}, demonstrates the mean relative error for the total mass, and hence lacks this problem. For a true $p^\star$ and an approximation $p$, defined on the $n \times n$ simulation grid, $1 \leq i, j \leq n$, the above error estimates read: 
\begin{align}
    \displaystyle
    \text{err}_1 &= \frac{1}{n^2} \ \sum_{i, j} \frac{|p^\star_{i, j} - p_{i, j}|} {p^\star_{i, j}}, \label{eq:err_1} \\ 
    \text{err}_2 &= \frac{\sum_{i, j} |p^\star_{i, j} - p_{i, j}|} {\sum_{i, j} p^\star_{i, j}} .    \label{eq:err_2}
\end{align}
The mean relative error  ranges from 0\% to 6\%.

\begin{figure}
    \centering
    \includegraphics[width=\linewidth]{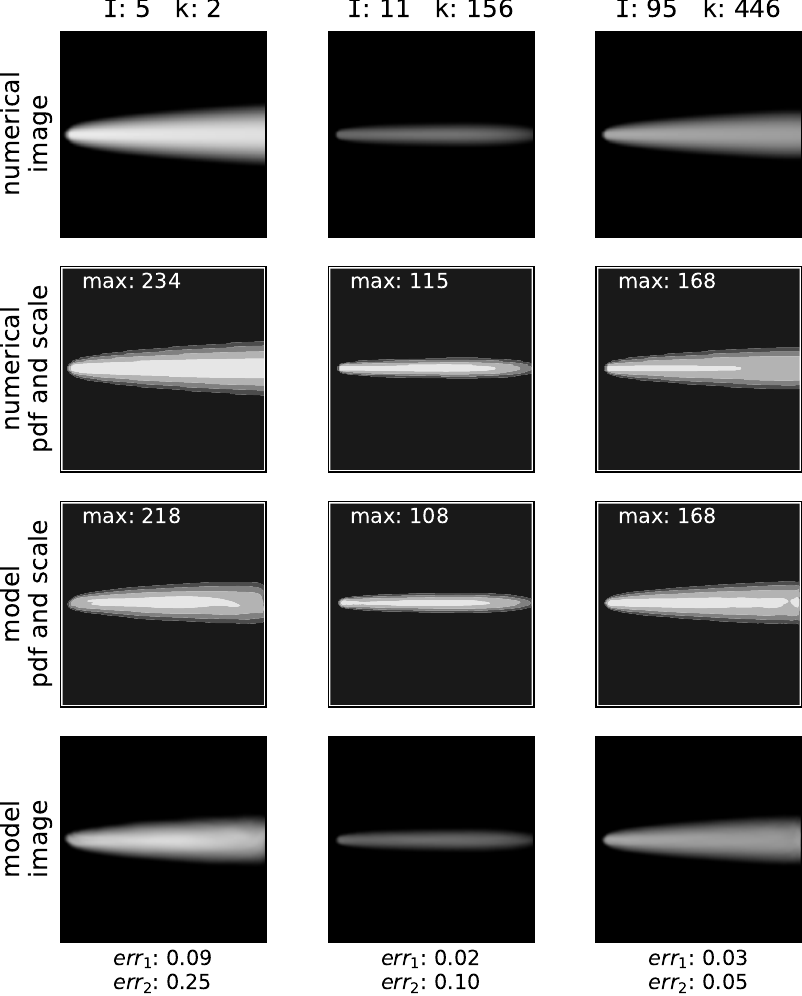}
    \caption{The prediction of unseen data images. The top row is the original images, the second row is their "technical" representations,  as course-grained $p(x, y| I, k)$ and brightness levels. The thirds row corresponds to the same representations, generated with the CNF and FFN modules and  the bottom row illustrates the reconstructed images. The mean relative error is about 10\%. }
    \label{fig:predicting_im}
\end{figure}

Fig. \ref{fig:predicting_im} illustrates the  prediction for unseen data. The mean relative error is higher, than for the seen data, and can attain 10\%.  Nevertheless, the result of the  overall data generation and reconstruction is rather satisfying and can be used for profiling of particles sizes. This is addressed in the next section.

\subsection{Spatial profiling of particles sizes}

In the reduced coordinate system, $(x, y) \in [-0.5, 0.5]$ we put the source of monomers to the $(-0.45, 0.0)$ point. To plot the  dependency of $\log c_k(x, y)$ as the function of the cluster size $k$ for particular point $(x,y)$, we place three probes in the simulation area:

\begin{enumerate}
    \item the orange one, near the source location at $(-0.4, 0.0)$
    \item the green one, in the center of the stream at $(0.1, 0.0)$
    \item the blue one, where density is decaying at $(0.3, 0.1)$
\end{enumerate}

\begin{figure*}[t]
    \centering
    \includegraphics[width=\linewidth]{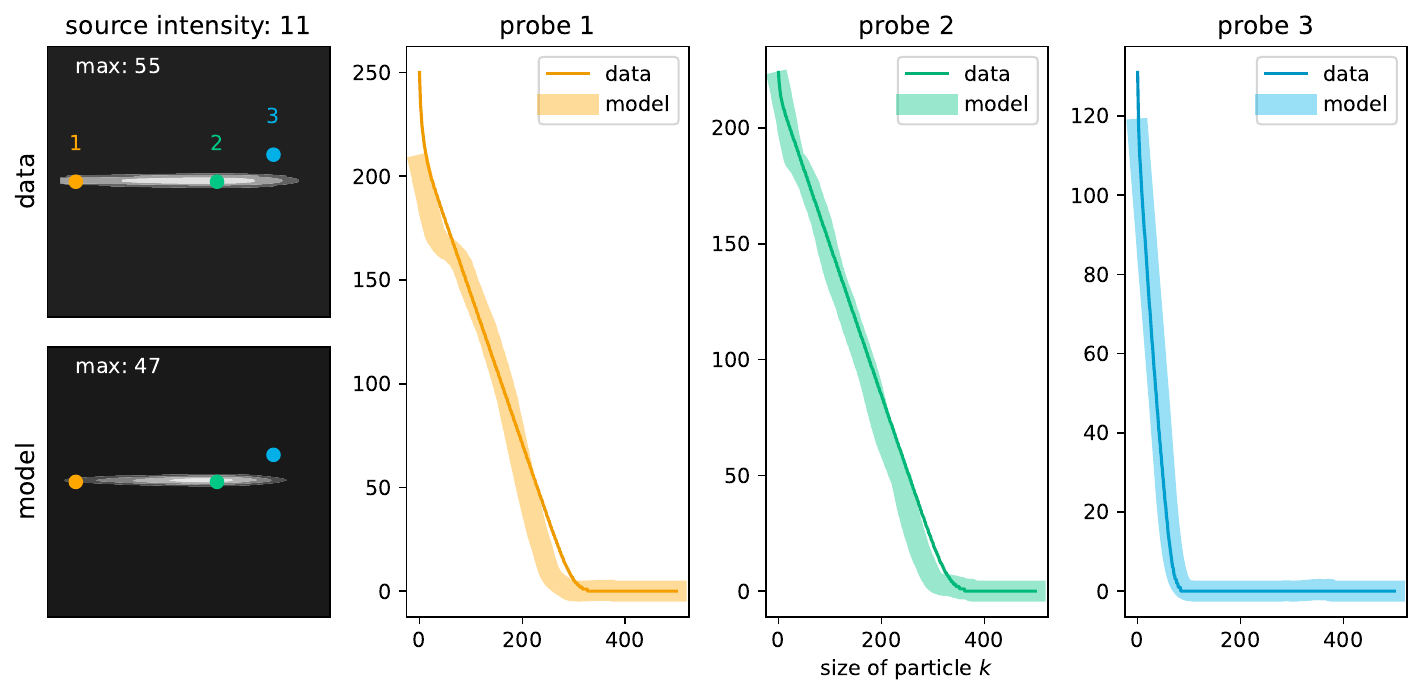}
    \includegraphics[width=\linewidth]{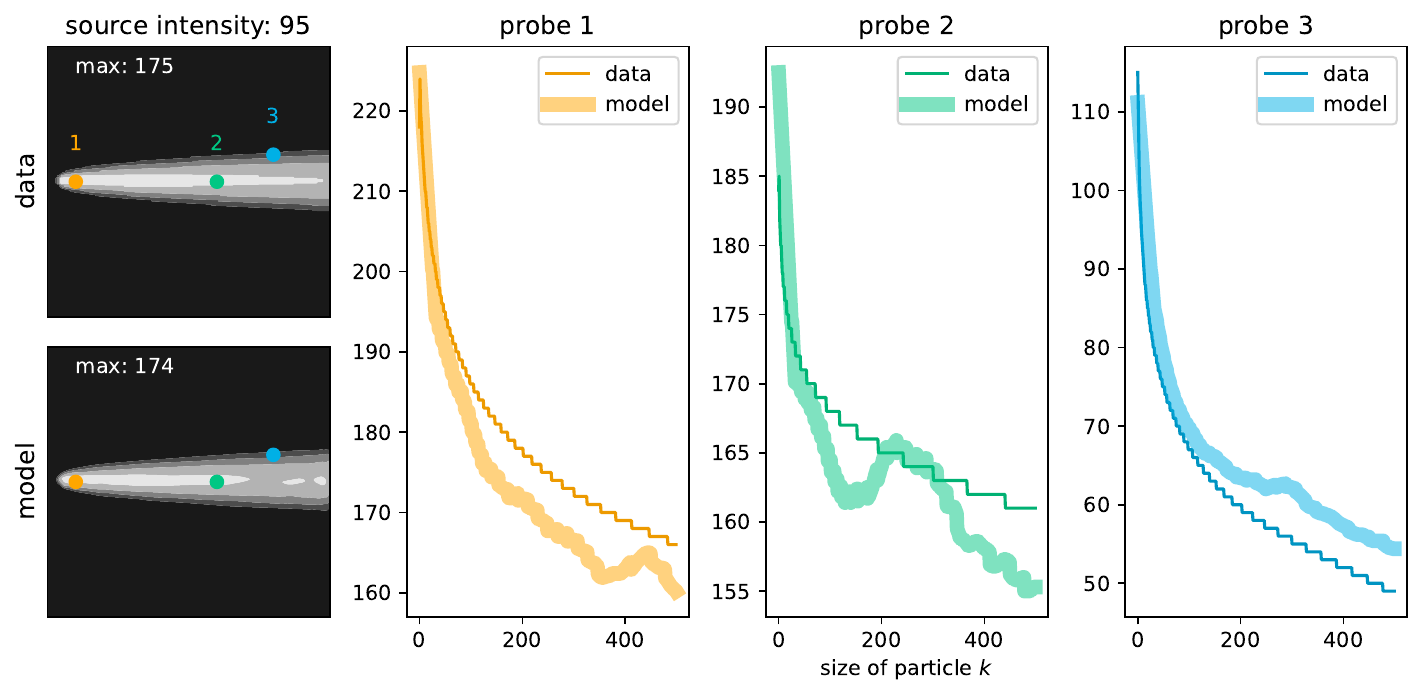}
    \caption{The spatial profiling of particles sizes. A comparison of the model predictions  and results of the numerical solution of the aggregation equations. Top panels: The low monomer source intensity; bottom panels: The high source intensity.}
    \label{fig:profiles}
\end{figure*}

With the use of the developed model the generation of the entire profile for $k_{\rm max}=500$ takes about 26 seconds, as  compared to approximately 1.5 hours for the numerical solution of the respective aggregation equations. The predicted profiles are in a good agreement with these obtained from the numerical solution, especially for the low intensity of monomer source, see Fig. \ref{fig:profiles}. Note, that the plots are given for the unseen data.

\section{Discussion and conclusion}

We explore the aggregation processes in two-dimensional space-inhomogeneous system, where the advection,  diffusion and a source of aggregating particles is present. To demonstrate the prove of concept, we restrict our study to the system, attaining a steady state and use a simplified model for the kinetic coefficients, namely, mass-independent aggregation rates and diffusion coefficients. We assume that the advection velocity is mass-independent as well. Firstly, we use the traditional approach for the problem and solve numerically  space-inhomogeneous Smoluchowski equations. To increase the computational efficiency, we apply here fast solvers, based on the low-rank decomposition of the aggregation kernel,  and the implicit scheme for the advection operator, to stabilise the numerical scheme for large P\'eclet numbers. We fix the values of the aggregation rates and diffusion coefficients and vary the intensity of the monomer source and the advection velocity. We observe that it suffices to consider only a reference value of the advection velocity, as the solution of all other values may be obtained by affine transformation of the coordinate system (rotation and re-scaling). Therefore we generated the training and test datasets, by solving the aggregation equations for the reference advection velocity and varying intensity of the monomer source. We form a training dataset of 10 different source intensity and test dataset of 7 intensities. 

Secondly, we apply the machine learning approach to predict the space-dependent size distribution of aggregating particles in systems with a monomer source. Namely, we build a generative model of relatively small size, suitable for our and similar, physics-related applications,  that require a high accuracy.  It consists of two parts: The conditional normalising flow module (with about $5.5\times10^5$ trainable parameters) and the fully connected feed-forward network (with about 10 000 weights). Hence, our model, containing about $5.6\times10^5$ parameters in total, is significantly more compact than nowadays models for image processing; for instance, the standard U-net model comprises $3\times10^7$ weights, see e.g.  \cite{ronneberger2015u}.

The ability to perform a direct density estimation makes the normalising flows model attractive, from the point of view of various applications. Indeed, the numerical simulations of many physical processes, often, naturally include various probability density functions (PDFs). In contrast to image segmentation, recognition and generation problems, hydrodynamics problems possess lower dimensions of PDFs, that is, they are two or three-dimensional.
Moreover, since the constructed model does not include any specific information about processes it approximates, the presented approach may be straightforwardly applied for similar problems as well.

\acknowledgments
The study was supported by a grant from the Russian Science Foundation No.~21-11-00363, https://rscf.ru/project/21-11-00363/.

\newpage

\bibliography{revtex_main}

\end{document}